\def\ie{{\it i.e.}}

\def\~{{$\tilde{\phantom{a}}$}}

\documentclass [12pt] {article}
\usepackage{epsfig}
\usepackage{color}

\def\thebibliography#1{\section{References}\markboth
 {REFERENCES}{REFERENCES}\list
 {[\arabic{enumi}]}{\settowidth\labelwidth{[#1]}\leftmargin\labelwidth
 \advance\leftmargin\labelsep
 \usecounter{enumi}}
 \def\newblock{\hskip .11em plus .33em minus -.07em}
 \sloppy
 \sfcode`\.=1000\relax}
\def\upcite#1{\raise6pt\hbox{\scriptsize
\cite{#1}}}
\def\lsim{\mathrel {\vcenter {\baselineskip 0pt \kern 0pt
    \hbox{$<$} \kern 0pt \hbox{$\sim$} }}}
\def\gsim{\mathrel {\vcenter {\baselineskip 0pt \kern 0pt
    \hbox{$>$} \kern 0pt \hbox{$\sim$} }}}
\def\gtlt{\mathrel {\vcenter {\baselineskip 0pt \kern 0pt
    \hbox{$>$} \kern 0pt \hbox{$<$} }}}

\def\hline{\noalign{\hrule \vskip2pt}}

\def\|{\ifmmode\Vert\else \char`\|\fi}
\ifx\oldzeta\undefined                          
  \let\oldzeta=\zeta                            
  \def\zzeta{{\raise 2pt\hbox{$\oldzeta$}}}     
  \let\zeta=\zzeta                              
\fi

\ifx\oldchi\undefined                           
  \let\oldchi=\chi                              
  \def\cchi{{\raise 2pt\hbox{$\oldchi$}}}       
  \let\chi=\cchi                                
\fi

\def\frac#1#2{{#1 \over #2}}

\def\half{\ifinner {\scriptstyle {1 \over 2}}
   \else {1 \over 2} \fi}

\def\abs#1{\left\vert#1\right\vert}	

\def\simge{\mathrel{%
   \rlap{\raise 0.511ex \hbox{$>$}}{\lower 0.511ex \hbox{$\sim$}}}}
\def\simle{\mathrel{
   \rlap{\raise 0.511ex \hbox{$<$}}{\lower 0.511ex \hbox{$\sim$}}}}

\def\buildchar#1#2#3{{\null\!                   
   \mathop#1\limits^{#2}_{#3}                   
   \!\null}}                                    
\def\overcirc#1{\buildchar{#1}{\circ}{}}

\def\slashchar#1{\setbox0=\hbox{$#1$}           
   \dimen0=\wd0                                 
   \setbox1=\hbox{/} \dimen1=\wd1               
   \ifdim\dimen0>\dimen1                        
      \rlap{\hbox to \dimen0{\hfil/\hfil}}      
      #1                                        
   \else                                        
      \rlap{\hbox to \dimen1{\hfil$#1$\hfil}}   
      /                                         
   \fi}                                         %

\def\subrightarrow#1{
  \setbox0=\hbox{
    $\displaystyle\mathop{}
    \limits_{#1}$}
  \dimen0=\wd0
  \advance \dimen0 by .5em
  \mathrel{
    \mathop{\hbox to \dimen0{\rightarrowfill}}
       \limits_{#1}}}                           

\def\overlay#1#2{\ifmmode%
\setbox0=\hbox{$#1$}%
\setbox1=\hbox to\wd0{\hss$#2$\hss}\else%
\setbox0=\hbox{#1}%
\setbox1=\hbox to\wd0{\hss#2\hss}\fi%
#1\hskip-\wd0\box1 }

\def\pmb#1{\leavevmode\setbox0=\hbox{#1}%
\kern-.02em\copy0\kern-\wd0
\kern.04em\copy0\kern-\wd0
\kern-.02em\raise.04em\box0 }

\def\vereq#1#2{\lower3pt\vbox{\baselineskip1.5pt \lineskip1.5pt
\ialign{$\m@th#1\hfill##\hfil$\crcr#2\crcr\sim\crcr}}}

\def\tensor#1{\protect\@ontopof{#1}{\leftrightarrow}{1.15}\mathord{\box2}}
\def\overstar#1{\protect\@ontopof{#1}{\ast}{1.15}\mathord{\box2}}
\def\overdots#1{\protect\@ontopof{#1}{\cdots}{1.0}\mathord{\box2}}
\def\overcirc#1{\protect\@ontopof{#1}{\circ}{1.2}\mathord{\box2}}
\def\loarrow#1{\protect\@ontopof{#1}{\leftarrow}{1.15}\mathord{\box2}}
\def\roarrow#1{\protect\@ontopof{#1}{\rightarrow}{1.15}\mathord{\box2}}

\def\@ontopof#1#2#3{%
{\mathchoice
{\@@ontopof{#1}{#2}{#3}\displaystyle\scriptstyle}%
{\@@ontopof{#1}{#2}{#3}\textstyle\scriptstyle}%
{\@@ontopof{#1}{#2}{#3}\scriptstyle\scriptscriptstyle}%
{\@@ontopof{#1}{#2}{#3}\scriptscriptstyle\scriptscriptstyle}%
}%
}

\def\@@ontopof#1#2#3#4#5{%
\setbox0=\hbox{$#4#1$}%
\setbox1=\hbox{$#5#2$}%
\setbox2=\hbox{}\ht2=\ht0 \dp2=\dp0 %
\ifdim\wd0>\wd1 %
\setbox1=\hbox to\wd0{\hss\box1\hss}%
\mathord{\rlap{\raise#3\ht0\box1}\box0}%
\else   %
\setbox1=\hbox to.9\wd1{\hss\box1\hss}%
\setbox0=\hbox to\wd1{\hss$#4\relax#1$\hss}%
\mathord{\rlap{\copy0}\raise#3\ht0\box1}%
\fi
}%

\def\lambdabar{\protect\@lambdabar}
\def\@lambdabar{%
\relax
\bgroup
\def\@tempa{\hbox{\raise.73\ht0
\hbox to0pt{\kern.25\wd0\vrule width.5\wd0
height.1pt depth.1pt\hss}\box0}}%
\mathchoice{\setbox0\hbox{$\displaystyle\lambda$}\@tempa}%
{\setbox0\hbox{$\textstyle\lambda$}\@tempa}%
{\setbox0\hbox{$\scriptstyle\lambda$}\@tempa}%
{\setbox0\hbox{$\scriptscriptstyle\lambda$}\@tempa}%
\egroup
}

\def\corresponds{{\lower.2ex\hbox{=}}{\rm\kern-.75em^\triangle}}
\def\succsim{\succ\kern-.9em_\sim\kern.3em}
\def\precsim{\prec\kern-1em_\sim\kern.3em}
\def\slantfrac#1#2{\kern1em^{#1}\kern-.3em/\kern-.1em_{#2}}

\begin{document}
                                                                
\begin{center}
{\Large\bf Magnetostatic Spin Waves}
\\

\medskip

Kirk T.~McDonald
\\
{\sl Joseph Henry Laboratories, Princeton University, Princeton, NJ 08544}
\\
(September 15, 2002)
\end{center}

\section{Problem}

Magnetostatics can be defined as the regime in which the magnetic fields
{\bf B} and {\bf H} have no time dependence, and ``of course'' the
electric fields {\bf D} and {\bf E} have no time dependence 
either.  In this case, the divergence of the fourth Maxwell
equation,
\begin{equation}
\nabla \times {\bf H} = {4 \pi \over c} {\bf J}_{\rm free} 
+ {1 \over c} {\partial {\bf D} \over \partial t}\, ,
\label{p1}
\end{equation}
(in Gaussian units) implies that 
\begin{equation}
\nabla \cdot {\bf J}_{\rm free} = 0,
\label{p2}
\end{equation}
\ie, that the free currents flow in closed loops.  Likewise, the time
derivative of the fourth Maxwell equation implies that ${\bf J}_{\rm free}$
has no time dependence in magnetostatics.

Often, magnetostatics is taken to be the situation in which $\nabla \cdot 
{\bf J}_{\rm free} = 0$ and {\bf D}, {\bf E} and ${\bf J}_{\rm free}$
have no time dependence, without explicit assumption that {\bf B} and
{\bf H} also have no time dependence.  Discuss the possibility of
waves of {\bf B} and {\bf H}, consistent with the latter definition of
magnetostatics \cite{bernstein}.

Consider two specific examples of ``magnetostatic'' waves
in which ${\bf J}_{\rm free} = 0$:
\begin{enumerate}
\item
Ferromagnetic spin waves in a medium subject to zero external field, but
which has a uniform static magnetization that is large compared to that of
the wave.  That is, ${\bf M} = M_0 \hat{\bf z} + {\bf m}e^{i ({\bf k} \cdot
{\bf r} - \omega t)}$,
where $m \ll M_0$.
 Here, the quantum mechanical exchange interaction is the dominant 
self interaction of the wave, which leads to an effective magnetic field
in the sample given by ${\bf B}_{\rm eff} = \alpha \nabla^2 {\bf m}$,
where $\alpha$ is a constant of the medium.
\item
Waves in a ferrite cylinder in a uniform external magnetic field
parallel to its axis,
supposing the spatial variation of the wave is slight, so the exchange
interaction may be ignored.  Again, the time-depedendent part of the
magnetization is assumed small compared to the static part.  Show that
the waves consist of transverse, magnetostatic fields that rotate with a
``resonant'' angular velocity about the axis. 

\end{enumerate}
In practice, the spin waves are usually excited by an external rf field, which
is to be neglected here.

\section{Solution}

\subsection{General Remarks}

In both definitions of magnetostatics the electric field {\bf E} has no 
time dependence, $\partial {\bf E} / \partial t = 0$, so the magnetic 
field {\bf B} obeys  $\partial^2 {\bf B} / \partial t^2 = 0$, 
as follows on taking the time derivative of Faraday's law, 
\begin{equation}
\nabla \times {\bf E} = - {1 \over c} {\partial {\bf B} \over \partial t}
\label{s1}
\end{equation}
(in Gaussian units).  
In principle, this is consistent with a magnetic field that varies
linearly with time, ${\bf B}({\bf r},t) = {\bf B}_0({\bf r}) + 
{\bf B}_1({\bf r})t$.  However,
this leads to arbitrarily large magnetic fields at early and late times, and
is excluded on physical grounds.   Hence, 
any magnetic field {\bf B} that coexists with only
static electric fields is also static.

There remains the possibility of a ``magnetostatic wave" in a magnetic 
medium that involves the magnetic field
${\bf H}_{\rm wave}$ and magnetization density ${\bf M}_{\rm wave}$
 which are related by
\begin{equation}
0 = {\bf B}_{\rm wave} = {\bf H}_{\rm wave} + 4 \pi {\bf M}_{\rm wave}.
\label{s2}
\end{equation}
If there are no free currents in the medium, and any electric field is static,
then the fourth Maxwell equation is simply
\begin{equation}
\nabla \times {\bf H} = 0,
\label{s3}
\end{equation}
which defines a subset of magnetostatic phenomena.

\subsection{Ferromagnetic Spin Waves}

Consider a ferromagnetic material that consists of a single macroscopic domain
with magnetization density ${\bf M} = M_0 \hat{\bf z} 
+ {\bf m}({\bf r},t)$, where $M_0$ is constant and $m \ll M_0$.  We
suppose there are no external electromagnetic fields.  Associated with
the magnetization {\bf M} are magnetic fields {\bf B} and {\bf H} whose
values depend on the geometry of the sample.  We suppose that the
weak time-dependent magnetic fields due to {\bf m} lead to even weaker
time-dependent electric fields, such that the situation is essentially
magnetostatic.  The consistency of this assumption will be confirmed at
the end of the analysis.

The ferromagnetism is due to electron spins, whose dominant interaction
is the quantum mechanical exchange interaction, in the absence of
external fields.  For a weak perturbation {\bf m} of the magnetization,
the exchange interaction preserves the magnitude of the magnetization,
so its time evolution has the form of a precession \cite{Landau},
\begin{equation}
{d {\bf M} \over dt} = \vec \Omega \times {\bf M}.
\label{s4}
\end{equation}
As this is the same form as the precession of a magnetic moment in an
external magnetic field \cite{magnetic-waves}, 
the precession vector $\vec\Omega$ is often
written as a gyromagnetic factor $\Gamma = e / 2 m_e c
\approx 10^7$ Hz/gauss times an effective magnetic 
field ${\bf B}_{\rm eff}$ (or ${\bf H}_{\rm eff}$).
Here, $e > 0$ and $m_e$ are the charge and mass of the electron, and $c$ is the
speed of light.  For a weak perturbation in an isotropic medium \cite{Landau},
\begin{equation}
{\bf B}_{\rm eff} 
= \alpha \nabla^2 {\bf m},
\label{s5}
\end{equation}
where $\alpha$ is a constant of the medium.

 Then, the equation of motion
of the magnetization {\bf m} is
\begin{equation}
{d {\bf m} \over dt} = \alpha \Gamma \nabla^2 {\bf m} \times {\bf M}.
\label{s6}
\end{equation}

For a plane-wave perturbation, whose phase factor is $e^{i ({\bf k} \cdot
{\bf r} - \omega t)}$, the equation of motion (\ref{s6}) becomes
\begin{equation}
i \omega {\bf m}  = \alpha \Gamma k^2 {\bf m} \times M_0 \hat{\bf z}.
\label{s7}
\end{equation}
This is satisfied by a circularly polarized wave,
\begin{equation}
{\bf m}  = m (\hat{\bf x} + i \hat{\bf y}) e^{i ({\bf k} \cdot
{\bf r} - \omega t)},
\label{s8}
\end{equation}
that obeys the quadratic dispersion relation \cite{Bloch}
\begin{equation}
\omega = \alpha \Gamma M_0 k^2,
\label{s9}
\end{equation}
which implies that $\omega \ll c k$ in physical materials, where $c$ is
the speed of light.  Hence, the electric fields are much smaller than the
magnetic fields associated with the time-dependent magnetization {\bf m},
so that $\nabla \times {\bf H} = 0$ to a good approximation, and we 
may use the term ``magnetostatic'' to describe the waves.  These
waves of magnetization are, however, better termed ``spin waves'', whose
quanta are called ``magnons''.

\subsection{Rotating Magnetostatic Modes in a Ferrite Cylinder}

In the magnetostatic approximation the fields {\bf B} and {\bf H} obey
\begin{equation}
\nabla \cdot {\bf B} = \nabla \cdot ({\bf H} + 4 \pi {\bf M}) = 0, \qquad
\nabla \times {\bf H} = 0,
\label{s21}
\end{equation}
where the field {\bf B} but not {\bf H} and {\bf M} must be static (or at least
so slowly varying in time that the resulting electric field is small compared
to {\bf B}).  We first
consider a ferrite of arbitrary shape of characteristic length $a$ in a uniform external magnetic field ${\bf B}_{\rm ext} = {\bf H}_{\rm ext} =
H_0 \hat{\bf z}$.  We suppose that this field is strong enough to induce
a uniform magnetization $M_0 \hat{\bf z}$ throughout the sample.

For waves with weak spatial dependence as we shall assume, the exchange
interaction is negligible, since it varies as the second spatial derivative 
of {\bf M}.  Then, the spins interact primarily with the local magnetic
field {\bf B} according to
\begin{equation}
{d {\bf M} \over d t} = \Gamma {\bf B} \times {\bf M}
= \Gamma {\bf H} \times {\bf M}.
\label{s22}
\end{equation}

We consider a perturbation {\bf m} to the magnetization that has
frequency $\omega$ and wavelength large compared to the size of the
the sample.   Then the total magnetization can be written
\begin{equation}
{\bf M} = M_0 \hat{\bf z} + {\bf m} e^{- i \omega t},
\label{s23}
\end{equation}
where $m \ll M_0$.  Similarly, we write the magnetic field inside the 
sample as
\begin{equation}
{\bf B} = B_z \hat{\bf z} + {\bf b} e^{- i \omega t}, \qquad
{\bf H} = H_z \hat{\bf z} + {\bf h} e^{- i \omega t},
\label{s24}
\end{equation}
where $B_z = H_z + 4 \pi M_0$ and $H_z = H_0 - 4 \pi N_z M_0$ are the sum 
of the external field and that due
to the uniform magnetization $M_0\hat{\bf z}$, and so are also uniform for
spheroidal (and cylindrical) samples whose axis is the $z$ axis \cite{Smythe}.
The ``demagnetization'' factor $N_z$ varies between 1 for a disk and 0 for a
cylinder. 
The perturbation {\bf m} exists only inside the sample, but the corresponding
perturbations {\bf b} and {\bf h} exist outside the sample as well.

Inserting eqs.~(\ref{s23}) and (\ref{s24}) in the equation of motion (\ref{s22}),
we keep only the first-order terms to find
\begin{equation}
- i \omega {\bf m} = \Gamma \hat{\bf z} \times (M_0 {\bf h} - H_z {\bf m}),
\label{s25}
\end{equation}
whose components are
\begin{eqnarray}
m_x & = & i {\Gamma \over \omega} (M_0 h_y - H_z m_y),
\nonumber \\
m_y & = & - i {\Gamma \over \omega} (M_0 h_x - H_z m_x),
\label{s26} \\
m_z & = & 0.
\nonumber \\
\end{eqnarray}
We solve for {\bf m} in terms of {\bf h} as
\begin{eqnarray}
m_x & = & \alpha h_x - i \beta h_y, 
\nonumber \\
m_y & = & i \beta h_x + \alpha h_y,
\label{s26a}
\end{eqnarray}
where
\begin{equation}
\alpha = {\Gamma^2 H_z M_0  \over \Gamma^2 H_z^2 - \omega^2}\, ,
\qquad
\beta = {\Gamma M_0 \omega  \over \Gamma^2 H_z^2 - \omega^2}\, .
\label{s26b}
\end{equation}
For later use, we note that in cylindrical coordinates, $(r,\theta,z)$, 
eq.~(\ref{s26a}) becomes
\begin{eqnarray}
m_r & = & \alpha h_r - i \beta h_\theta, 
\nonumber \\
m_\theta & = & i \beta h_r + \alpha h_\theta.
\label{s26c}
\end{eqnarray}

As we are working in the magnetostatic limit (\ref{s21}), we also have
\begin{equation}
\nabla \cdot {\bf b} = \nabla \cdot ({\bf h} + 4 \pi {\bf m}) = 0, \qquad
\nabla \times {\bf h} = 0.
\label{s27}
\end{equation}
Hence, the perturbation {\bf h} can be derived from a scalar potential,
\begin{equation}
{\bf h} = - \nabla \phi,
\label{s28}
\end{equation}
and so,
\begin{equation}
\nabla^2 \phi = 4 \pi \nabla \cdot {\bf m}.
\label{s29}
\end{equation}
Outside the sample the potential obeys Laplace's equation,
\begin{equation}
\nabla^2 \phi = 0 \qquad \mbox{(outside)},
\label{s30}
\end{equation}
while inside the sample we find, using eq.~(\ref{s26a}),
\begin{equation}
(1 + 4 \pi \alpha) \left( {\partial^2 \phi \over \partial x^2}
+ {\partial^2 \phi \over \partial y^2} \right) 
+ {\partial^2 \phi \over \partial z^2} = 0 \qquad \mbox{(inside)}.
\label{s31}
\end{equation}

The case of an oblate or prolate spheroid with axis along the external field
has been solved with great virtuosity by Walker \cite{Walker}, following the
realization that higher-order modes deserved discussion \cite{Feynman}.  Here, we
content ourselves with the much simpler case of a long cylinder whose axis is
along the external field, for which the lowest-order spatial mode was first
discussed by Kittel \cite{Kittel}.
We consider only the case of waves with no spatial dependence
along the axis of the cylinder.

With these restrictions, both eqs.~(\ref{s30}) and (\ref{s31}) reduce to
Laplace's equation in two dimensions.  We can now work in a cylindrical
coordinate system $(r,\theta,z)$, where appropriate 2-D solutions to
Laplace's equation have the form
\begin{eqnarray}
\phi(r<a,\theta) & = & \sum {r^n \over a^n} 
(A_n e^{i n \theta} + B_n e^{-i n \theta}),
\label{s32} \\
\phi(r>a,\theta) & = & \sum {a^n \over r^n}
(A_n e^{i n \theta} + B_n e^{-i n \theta}),
\label{s33}
\end{eqnarray}
which is finite at $r = 0$ and $\infty$, has period $2 \pi$ in $\theta$,
and is continuous at the boundary $r = a$.

The boundary conditions at $r = a$ in the magnetostatic limit (\ref{s27})
are that $b_r$ and $h_\theta$ are continuous.  The latter condition is
already satisfied, since $h_\theta = -(1 / r) \partial \phi / \partial \theta$.
We note that
\begin{equation}
b_r = h_r + 4 \pi m_r  
= (1 + 4 \pi \alpha) h_r - 4 \pi i \beta h_\theta,
\label{s34}
\end{equation}
recalling eq.~(\ref{s26c}).  Using eqs.~(\ref{s32}) and (\ref{s33}) we find
that continuity of $b_r$ at $ r = a$ requires
\begin{equation}
\sum {n \over a} \left[ (1 + 2 \pi \alpha + 2 \pi \beta) A_n e^{i n \theta} + 
(1 + 2 \pi \alpha - 2 \pi \beta) B_n e^{-i n \theta} \right] = 0.
\label{s35}
\end{equation}
Nontrivial solutions are possible only if $2 \pi (\alpha \pm \beta) = -1$, 
in either of which case there is an infinite
set of modes that are degenerate in frequency.  
Using eq.~(\ref{s26b}), we find the ``resonance'' frequency to be
\begin{equation}
\omega = \pm \Gamma (H_0 + 2 \pi M_0),
\label{s36}
\end{equation}
noting that for a cylinder the demagnetization factor is $N_z = 0$, so that
$H_z = H_0$, as is readily deduced by elementary arguments.  Since we
consider frequency $\omega$ to be positive, we see that the two solutions
(\ref{s36}) correspond to two signs of $H_0$, and are essentially identical.

For spheroidal samples, the modes are enumerated with two integer indices, and
are not all degenerate in frequency, as discussed in \cite{Walker}.

We close our discussion by showing that the electric field of the wave is
much smaller than the magnetic field.  The scalar potential for mode $n$ is
\begin{equation}
\phi_n(r < a) = {r^n \over a^n} e^{i(n \theta - \omega t)}, \qquad
\phi_n(r > a) = {a^n \over r^n} e^{i(n \theta - \omega t)}.
\label{s37}
\end{equation}
We see that for $n > 0$ the potential rotates with angular velocity $\Omega_n
= \omega / n$ about the $z$ axis.  The potential is maximal at $r = a$, so
consistency with special relativity requires that 
\begin{equation}
v(r=a) = {a \omega \over n} \ll c,
\label{s37a}
\end{equation}
which appears to have been (barely) satisfied in typical experiments
\cite{Kittel}.  We also see that for high mode number the spatial variation
of the wave becomes rapid, and the neglect of the exchange interaction is
no longer justified.

The magnetic field ${\bf h} = - \nabla \phi$ of mode $n$ has components
\begin{equation}
h_r(r < a) = - n{r^{n-1} \over a^n} e^{i(n \theta - \omega t)}
= - {n \over r} \phi_n, \qquad
h_r(r > a) = n {a^n \over r^{n+1}} e^{i(n \theta - \omega t)}
= {n \over r} \phi_n,
\label{s38}
\end{equation}
\begin{equation}
h_\theta(r < a) = i h_r(r < a), \qquad h_\theta(r > a) = - i h_r(r > a).
\label{s39}
\end{equation}
The monopole mode, $n = 0$, does not exist.  The lowest mode is $ n = 1$,
which corresponds to a uniform, transverse field {\bf h} that rotates about
the $z$ axis with angular velocity $\omega$.

From eq.~(\ref{s26c}) we find the magnetization to be
\begin{equation}
{\bf m} = - {{\bf h} \over 2 \pi}
\label{s39a}
\end{equation}
for all modes (for $r < a$ only, of course), so the magnetization of mode $n$
also rotates with angular velocity $\omega / n$.

The magnetic field ${\bf b} = {\bf h} + 4 \pi {\bf m}$ is then,
\begin{equation}
{\bf b}(r < a) = - {\bf h}(r < a),  \qquad
{\bf b}(r > a) = {\bf h}(r > a).
\label{s40}
\end{equation}
Using either the $r$ or $\theta$
component of Faraday's law, we find that the associated electric field 
{\bf e} has only a $z$ component,
\begin{equation}
{\bf e} = {\omega \over c} \phi \hat{\bf z}\, ,
\label{s42}
\end{equation}
both inside and outside the cylinder (consistent with continuity of the
tangential component of the electric field at a boundary, and with $\nabla
\cdot {\bf e} = 0$).
The ratio of the electric to the magnetic field of mode $n$ at $r = a$ is
\begin{equation}
\abs{e_z \over b_r} = {a \omega \over n c}\, ,
\label{s43}
\end{equation}
which is small so long as condition (\ref{s37a}) is satisfied.  Hence,
the condition that $a \omega / c \ll 1$ is doubly necessary for the 
validity of this analysis.

Because the magnetization {\bf m} is moving (rotating), there is an
associated electric polarization {\bf p} according to special relativity
\cite{Becker}, 
\begin{equation}
{\bf p} = \gamma {{\bf v} \over c} \times {\bf m}.
\label{s44}
\end{equation}
For mode $n$ we have ${\bf v} = \omega r \hat\theta / n \ll c$, so $\gamma
= 1 / \sqrt{1 - v^2 / c^2} \approx 1$, and
\begin{equation}
{\bf p} = - {\omega r m_r \over n c} \hat{\bf z}
= {\omega r h_r \over 2 \pi n c} \hat{\bf z}
= - {\omega \over 2 \pi c} \phi\ \hat{\bf z}
= - {{\bf e} \over 2 \pi}\, .
\label{s45}
\end{equation}

The electric displacement is related by ${\bf d} = {\bf e} + 4 \pi {\bf p}$,
which has the value
\begin{equation}
{\bf d}(r < a) = - {\bf e} = - {\omega \over c} \phi\ \hat{\bf z}\, ,
\qquad
{\bf d}(r > a) = {\bf e} = {\omega \over c} \phi\ \hat{\bf z}\, ,
\label{s46}
\end{equation}
The fourth Maxwell equation now implies that
\begin{equation}
\nabla \times {\bf h} = - \nabla \times \nabla \phi 
= {1 \over c} {\partial {\bf d} \over \partial t}
= \pm i {\omega^2 \over c^2} \phi\ \hat{\bf z}
\label{s47}
\end{equation}
Thus, there is a small violation of the magnetostatic conditions (\ref{s27}),
but this is second order in the small quantity $a \omega / c$ (noting that
$\nabla \phi \approx \phi / a$).

\end{document}